\documentclass{article}

\usepackage{arxiv}

\usepackage[utf8]{inputenc} 
\usepackage[T1]{fontenc}    
\usepackage{hyperref}       
\usepackage{url}            
\usepackage{booktabs}       
\usepackage{amsfonts}       
\usepackage{nicefrac}       
\usepackage{microtype}      
\usepackage{lipsum}		
\usepackage{graphicx}
\usepackage{natbib}
\usepackage{doi}

\title{LLMs Have Made Failure Worth Publishing}

\author{
Sungmin Lee \\
\texttt{i.am.sungmin.lee@gmail.com}
}   

\renewcommand{\shorttitle}{\textit{arXiv} Template}

\hypersetup{
pdftitle={LLMs May Have Made Failure Worth Publishing},
pdfauthor={},
}

\begin{document}
\maketitle

\begin{abstract}
Scientific publishing systematically filters out negative results.
We argue that this long-standing asymmetry has become an urgent
problem in the era of large language models, which inherit the
positive bias of the literature they are trained on, face an
impending shortage of high-quality training data, and are
increasingly deployed as both research tools and peer reviewers. We
analyze three ways in which LLMs have changed the value of failure
data and show that the systematic absence of such data degrades
their utility as research tools, training data consumers, and peer
reviewers alike. We outline experimental protocols to validate
these claims and discuss the structural conditions under which a
failure-inclusive publishing culture could emerge.
\end{abstract}

\section{Introduction}
\label{sec:intro}

Scientific publishing filters for success. Positive results
dominate the published literature across disciplines
\citep{fanelli2012negative}, and the pattern persists not because
scientists ask better questions but because negative findings are
systematically filtered out. Franco et al. found that roughly
two-thirds of null results from NSF-funded experiments were never
even written up \citep{franco2014publication}. The body of
scientific effort lost through this process is what Rosenthal
called the ``file drawer'' \citep{rosenthal1979file}.

The file drawer was not a failure of the system but an inevitable
triage under limited human retrieval and digestion bandwidth. When
the cost of finding a relevant failure exceeded the cost of simply
trying again, communities naturally prioritized sharing what worked
over documenting what did not.

This article argues that the arrival of large language models has
fundamentally altered this calculus, making the systematic
publishing of failure both feasible and urgent. We review the scale
and growing costs of the file drawer
(Section~\ref{sec:background}), analyze three ways in which LLMs
have changed the value of failure data
(Section~\ref{sec:opportunity}), propose experimental protocols for
validating these claims (Section~\ref{sec:future}), and discuss the
conditions for responsible use (Section~\ref{sec:discussion}).
\section{The File Drawer and Its Growing Costs}
\label{sec:background}

\subsection{The Scale of Publication Bias}

The extent of positive filtering has been quantified across multiple
disciplines. Fanelli analyzed over 4,600 papers published between
1990 and 2007 and found that the proportion reporting positive
results exceeded 80\% after 1999, peaking at 88.6\% in 2005
\citep{fanelli2012negative}. In one subfield where this pattern has
been directly quantified, McGreivy and Hakim reviewed 82 articles on
ML methods for solving fluid-related PDEs and found that 93\% claimed
superiority over traditional numerical methods, yet 79\% of those
claims relied on weak baselines. The authors concluded that negative
results were absent not because ML almost always outperforms, but
because researchers almost never publish when it does not
\citep{mcgreivy2024replacing}. Scheel et al. provided a controlled
test by comparing standard psychology papers with Registered Reports,
a format in which publication is guaranteed regardless of outcome.
The positive result rate dropped from 96\% (N=152) to 44\% (N=71)
\citep{scheel2021excess}.

\subsection{Why Failure Matters}

Research fails far more often than it succeeds. Over 90\% of drug
candidates entering clinical trials fail to reach approval
\citep{sun2022clinical}. Most hypotheses in basic science do not
survive experimental testing. The knowledge gained from these
failures is substantial. Experienced researchers accumulate failed
attempts as tacit knowledge, building an internal distribution of
prior experience that sharpens judgment about which directions are
promising and which are dead ends. This is what makes a senior
scientist valuable, and what makes doctoral training long and
difficult. Years of absorbing unpublished failure, at significant
personal cost, gradually build the intuition that no textbook can
substitute.

\subsection{The Collective Cost of Not Sharing Failure}

At the collective level, this tacit knowledge transfers informally
through mentoring and conversation, but not through the published
literature, which records only what works. Every researcher has
experienced the moment of discovering, months or years into a
project, that someone elsewhere had already attempted the same
approach and quietly abandoned it. In published literature, this
appears as being ``scooped.'' But when the prior work failed and
was never published, there is no record to be scooped by, and the
redundancy remains invisible. Chalmers and Glasziou estimated that
85\% of total research investment is avoidably wasted, with failure
to account for prior evidence, including unpublished negative
results, identified as a primary driver
\citep{chalmers2009avoidable}.

These costs have been compounding. Over the past two decades,
research output has expanded steadily, multiplying both the volume
of redundant failure and the pressure on individual researchers to
produce publishable results. At the extreme end of this pressure,
questionable research practices, from selective reporting to data
manipulation, have become more prevalent in more competitive
academic environments \citep{fanelli2010pressures}. Suspected paper
mill articles are now doubling every 1.5 years, roughly ten times
faster than the total publication base \citep{richardson2025pnas}.
In 2023 alone, over 10,000 papers were retracted, and retraction
rates have more than tripled over the past decade
\citep{noorden2023retraction}. Yet the arrangement persisted,
because recording and retrieving failure at scale remained
impractical at human reading speed, and no alternative was
economical enough to justify the shift.
\section{How LLMs Change the Value of Failure}
\label{sec:opportunity}

The arrival of large language models has changed this calculus in
two directions simultaneously. On one hand, LLMs have lowered the
barrier to producing manuscripts, accelerating the very pressures
described above. The volume of submissions to major conferences
reflects this. ICLR received roughly 7,000 submissions in 2024,
12,000 in 2025, and nearly 20,000 in 2026
\citep{pangram2025iclr}. On the other hand, LLMs have made it
practical for the first time to publish, retrieve, and exploit
failure at scale. This section examines three ways in which this
changes the value of failure data.

\subsection{LLMs Have Relaxed the Retrieval Bottleneck}

A model processing over a hundred thousand words in seconds can
retrieve, compare, and synthesize recorded knowledge at a scale no
human reader ever could. This directly addresses the economic logic
that justified the file drawer. If failures were systematically
published, LLMs could surface them before a researcher commits
months to a direction that others have already explored and
abandoned. The cost of retrieval, which once exceeded the cost of
simply trying again, has dropped by orders of magnitude. The triage
that produced the file drawer is no longer necessary.

At the collective level, this shift compounds. The redundant
failures described in Section~\ref{sec:background}, invisible
because no individual researcher could track what every other group
had tried, become visible when an LLM can cross-reference failure
records across labs, disciplines, and languages simultaneously.
What was an unavoidable collective tax on scientific progress
becomes an avoidable inefficiency.

\subsection{LLMs Need Less Biased Training Data}

From a training perspective, LLMs face a data crisis. Frontier
models have consumed most available high-quality text
\citep{villalobos2024run}, and training on model-generated text
leads to distributional collapse \citep{shumailov2024curse}. The
field needs new, human-generated data. The largest untapped corpus
meeting this criterion is precisely the one that was never
published.

Beyond volume, the composition of training data matters. An LLM
trained predominantly on positive-result literature inherits a
distorted model of science in which most experiments succeed. This
positive skew may systematically miscalibrate the model's judgment
about which research directions are viable. Incorporating published
failure data would not only expand the training corpus but also
correct the distributional bias at its source.

Early evidence from domain-specific applications supports this.
\citet{toniato2025negative} incorporated failed chemical reactions
into language model training and found improved prediction accuracy,
demonstrating that negative reactions carry learnable signal because
each attempt is grounded in background knowledge.
\citet{lee2025banel} went further, showing that a generative model
can be improved using only failed attempts, with no positive
examples at all, achieving orders-of-magnitude gains in success
rate on sparse-reward tasks. \citet{naser2025failure} surveyed this
emerging direction more broadly, documenting techniques such as
negative knowledge distillation and error-based curriculum learning
that are designed to extract signal from suboptimal outcomes. These
results indicate that the methodological infrastructure to exploit
failure data already exists. What is missing is not the ability to
learn from failure but the published failure data to learn from.

\subsection{LLM Reviewers Need Failure to Judge Well}

The peer review system is already straining. At ICLR 2026, 21\% of
75,800 peer reviews were flagged as entirely AI-generated
\citep{pangram2025iclr}. At ICML 2025, papers were found to
contain hidden prompts designed to manipulate LLM reviewers into
giving favorable scores \citep{theocharopoulos2025prompt}. Multiple
major venues have responded with escalating bans and detection
measures in their calls for papers. But if the volume of
submissions continues to grow at the current rate, human-only
review will not scale. The question is not whether LLMs will
participate in peer review, but whether they will do so with the
judgment necessary to catch flawed work.

A reviewer, human or machine, trained exclusively on a literature
in which positive results exceed 85\%
\citep{fanelli2012negative} may lack a well-calibrated model of
what failure looks like. We hypothesize that exposure to structured
failure data would improve the ability of LLM reviewers to detect
methodological errors and implausible claims. This remains
untested, and we propose a direct experiment in
Section~\ref{sec:future}.

Current responses are predominantly defensive, deploying detection
tools in an escalating arms race against increasingly sophisticated
fabrication. A complementary approach is structural. Rather than
attempting to prevent bad science from entering the system, building
legitimate channels through which negative results can be published
and absorbed would equip both human and machine reviewers with a
more accurate model of scientific reality. The goal is not to block
the production of waste but to eliminate the conditions that make
waste the rational output.
\section{Proposed Experimental Validation}
\label{sec:future}

The central claims of this paper can be reduced to two testable
propositions. First, that LLMs inherit the positive bias of the
scientific literature. Second, that structured failure data corrects
this bias. We outline one experiment for each.

\subsection{Experiment 1. Do LLMs Overestimate Success?}

\textbf{Hypothesis.} LLMs systematically overestimate the
probability of experimental success due to positive publication bias
in their training data.

\textbf{Design.} Select 200--500 completed clinical trials from
ClinicalTrials.gov. For each trial, extract the information available
at the start of the trial (hypothesis, drug, target disease, phase,
design) and withhold the actual outcome. Prompt multiple LLMs to
estimate the probability of success based on the starting information
alone. Compare LLM predictions against actual outcomes.

\textbf{Key measures.} Overall AUC-ROC. Distribution of predicted
success probabilities for trials that actually failed. Comparison
across trial phases.

\textbf{Expected finding.} LLMs will assign systematically higher
success probabilities to trials that ultimately failed, reflecting
the positive skew of the literature they were trained on.

\subsection{Experiment 2. Does Failure Data Help?}

\textbf{Hypothesis.} Incorporating failure data into LLM context
improves predictive accuracy, but only when failure data is properly
classified.

\textbf{Design.} Using the same clinical trial dataset, compare LLM
performance under five conditions.

\begin{table}[h]
\centering
\begin{tabular}{ll}
\toprule
\textbf{Condition} & \textbf{Context provided} \\
\midrule
A (baseline)     & No additional context \\
B (success only) & Related successful trials only \\
C (failure only) & Related failed trials only \\
D (both)         & Related successful and failed trials \\
E (classified)   & D + failure cause analysis (Type A/B/C) \\
\bottomrule
\end{tabular}
\end{table}

\textbf{Key measures.} Prediction AUC-ROC per condition.
Dose-response curve (10, 50, 100, 200 failure records).

\textbf{Expected finding.} Condition B (success only) will
\textit{worsen} performance relative to baseline A, by reinforcing
positive bias. Condition D will outperform C, and E will outperform
D. If unclassified failure data degrades performance relative to
classified data, this validates the necessity of the proposed
taxonomy.

\section{Discussion}
\label{sec:discussion}

\subsection{Not All Failures Are Equal}

If failure data is to be published and used, a natural question
arises about whether all failures carry equal informational value.
We believe they do not, and that failing to distinguish between
types of failure risks degrading rather than improving any system
that consumes the data.

Consider a rejected paper. It may have been rejected because its
methodology was flawed, or because its hypothesis was genuinely
unsupported by a well-designed experiment. The former teaches us
\textit{how not to do science}. The latter teaches us \textit{what
nature does not permit}. A third category exists where methodology
and outcome are entangled, making it unclear whether the null
finding reflects a true absence of effect or a flaw in execution.

We tentatively propose a three-part classification. Type A
(methodological failure) captures cases where the approach itself
was flawed, such as applying an inappropriate statistical test,
using an underpowered sample, or failing to control for known
confounders. Type B (substantive null result) captures cases where
a sound experiment yielded a negative outcome, such as a
well-powered clinical trial that finds no effect or a carefully
controlled replication that fails to reproduce a prior finding.
Type C (ambiguous) captures cases where the two cannot be
separated, as when a reviewer notes that it is unclear whether the
null result reflects a true absence of effect or inadequate
execution.

\subsection{What Counts as a Meaningful Failure Publication}

This paper is not an argument for publishing every unsuccessful
attempt or minor negative variation. An uncurated flood of
low-effort failure reports would introduce its own form of noise,
potentially worse than the current silence. The value of failure
data depends entirely on how it is defined, structured, and
quality-controlled.

What constitutes a meaningful failure publication will differ across
disciplines. A well-powered clinical trial that finds no effect is
qualitatively different from a preliminary computational experiment
that did not converge. The former represents substantial investment
and produces definitive evidence about a hypothesis. The latter may
reflect an implementation choice rather than a scientific finding.
Each field will need to develop its own criteria for what meets the
threshold of a publishable failure, just as each field has developed
its own standards for what meets the threshold of a publishable
success. This process of definition and community acceptance is a
prerequisite for any failure publication infrastructure to function,
and cannot be imposed top-down.

For instance, a substantive null result from a well-designed
experiment (Type B) is almost always informative. A methodological
failure (Type A) is informative only when the error is common
enough that documenting it would prevent others from repeating it.
The threshold for publication will naturally differ between these
categories.

\subsection{Further Research Directions}

Beyond the two core experiments proposed in
Section~\ref{sec:future}, several additional lines of investigation
would strengthen the case presented here.

One natural extension is to test whether the relationship between
publication bias and LLM error holds across fields. If the published
positive-result ratio and the actual success ratio (obtainable from
trial registries) were measured independently for multiple medical
subfields, a positive correlation between a field's publication
bias index and LLM prediction error would provide mechanistic
evidence that publication bias is a direct cause of LLM
miscalibration, not merely a co-occurring phenomenon.

Another direction concerns LLM-assisted peer review. As discussed
in Section~\ref{sec:opportunity}, LLM reviewers trained on
positively-biased literature may lack the distributional knowledge
to identify flawed work. An experiment comparing LLM review quality
with and without failure data in the context, using publicly
available reviews from OpenReview, would directly test whether
failure exposure improves the ability to detect methodological
errors and reduces hallucinated praise. Given that 21\% of ICLR
2026 reviews were flagged as AI-generated
\citep{pangram2025iclr}, the practical relevance of this question
is immediate.

\subsection{Structural versus Defensive Responses}

Current responses to the publishing and review crisis, including
LLM detection tools, usage bans, and prompt-injection
countermeasures, are necessary but insufficient. They are defensive
measures in an arms race that the detection side is unlikely to win
permanently. A complementary approach is to change the structural
conditions that make gaming the system rational. If failure had a
legitimate place in the scientific record, the incentive to disguise
it as success would diminish, and the tools trained on that record
would develop a more accurate model of scientific reality.

We emphasize that the connection between failure publication and
fraud reduction is structural, not causal. It is plausible that
legitimizing failure as a recognized scholarly output would reduce
the pressure to fabricate. However, unless hiring and promotion
committees recognize failure contributions as scholarly merit, the
underlying pressure remains unchanged regardless of whether
publishing infrastructure exists. The causal verification of this
link requires longitudinal intervention studies that are beyond the
scope of this work.

\subsection{Limitations}

This paper presents a conceptual framework and does not include
empirical validation. While the individual observations we draw
upon are well-established, the central claim that these converge
into a unified case for failure publication infrastructure remains
an argument, not a demonstrated fact.

Several specific limitations should be noted. First, our argument
that failure data improves LLM performance rests partly on analogy
with human expertise. Senior scientists benefit from accumulated
failure experience, but it does not automatically follow that the
same information, encoded as text, will produce comparable
improvements in a statistical language model. The mechanisms of
human intuition and LLM pattern matching are fundamentally
different, and the transferability of failure information across
these systems requires direct experimental verification.

Second, we do not address the practical challenges of incentivizing
failure publication at scale, including questions of intellectual
property, competitive risk, and the additional labor burden on
researchers who are already under pressure. A failure publication
system that no one uses solves nothing.

Third, the proposed failure taxonomy is preliminary and untested.
Its value depends on whether the categories can be applied
consistently and whether the distinction actually affects downstream
LLM performance, both of which are empirical questions.
\section{Conclusion}
\label{sec:conclusion}

The file drawer was a rational adaptation to a world in which
retrieving failure was more expensive than repeating it. Large
language models have inverted this cost structure. Failure data that
was once too costly to retrieve and too voluminous to digest can now
be searched, synthesized, and applied at scale. The published
scientific record is positively biased, and the systems learning
from it inherit that bias. The technology to change this now exists.
What remains is the harder problem of defining what a meaningful
failure publication looks like in each field, building the
infrastructure to support it, and creating the institutional
incentives that make contributing to it worthwhile.

\bibliographystyle{unsrtnat}
\bibliography{references}  






\end{document}